\documentclass[journal=jacsat,manuscript=article]{achemso}

\usepackage[version=3]{mhchem} 
\usepackage{textgreek}

\author{Daniel P. Arnold}
\affiliation[UCSB]
{Department of Chemical Engineering, University of California, Santa Barbara, Santa Barbara, CA, 93106}
\author{Sho C. Takatori}
\email{stakatori@ucsb.edu}
\affiliation[UCSB]
{Department of Chemical Engineering, University of California, Santa Barbara, Santa Barbara, CA, 93106}

\title{Lipid membrane domains control actin network viscoelasticity}

\begin{document}

\begin{tocentry}

\includegraphics[width=\linewidth]{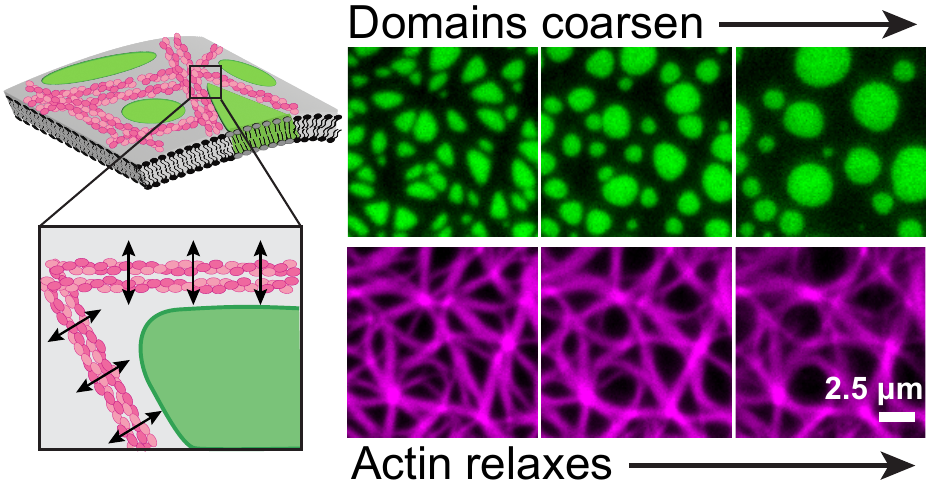}

\end{tocentry}

\begin{abstract}
The mammalian cell membrane is embedded with biomolecular condensates of protein and lipid clusters, which interact with an underlying viscoelastic cytoskeleton network to organize the cell surface and mechanically interact with the extracellular environment.
However, the mechanical and thermodynamic interplay between the viscoelastic network and liquid-liquid phase separation of 2-dimensional (2D) lipid condensates remains poorly understood.
Here, we engineer materials composed of 2D lipid membrane condensates embedded within a thin viscoelastic actin network. 
The network generates localized anisotropic stresses that deform lipid condensates into triangular morphologies with sharp edges and corners, shapes unseen in 3D composite gels.
Kinetic coarsening of phase-separating lipid condensates accelerates the viscoelastic relaxation of the network, leading to an effectively softer composite material over intermediate timescales.
We dynamically manipulate the membrane composition to control the elastic-to-viscous crossover of the network.
Such viscoelastic composite membranes may enable the development of coatings, catalytic surfaces, separation membranes, and other interfaces with tunable spatial organization and plasticity mechanisms.
\end{abstract}

\section{Introduction}

Lipid bilayer membranes are 2D interfacial materials that elastically resist bending and stretching, but exhibit viscous in-plane behavior \cite{Israelachvili2010}.
In the mammalian cell, a network of semiflexible actin filaments underlies and applies forces to the plasma membrane, allowing the cell to mechanically interact with its environment \cite{Alberts2017, Phillips2009}.
In addition to generating forces orthogonal to the membrane, the cytoskeleton also plays a role in organizing the spatial distribution and movement of protein and lipid clusters on the cell surface \cite{Kusumi1996, Fujiwara2016, Viola2007}.

Reconstituted lipid membranes are often used to study the interactions between heterogeneous membranes and the viscoelastic actin cytoskeleton \cite{Arnold2023a, Arnold2023b, Gubbala2024, Honigmann2014, Vogel2017}.
Multicomponent lipid membranes can undergo 2D liquid-liquid phase separation at room temperature \cite{Veatch2003}, and lipid compositions can be tuned to make actin selectively adsorb onto a single phase of the membrane \cite{Arnold2023a, Arnold2023b, Honigmann2014, Vogel2017, Schroer2020}.
A dense 2D concentration of filamentous actin can form viscoelastic networks on the membrane that alter the lipid phase behavior \cite{Arnold2023b}, much like 3D elastic polymer gels, which alter the growth of embedded liquid inclusions \cite{Style2018,Rosowski2020,Rosowski2020a,Fernandez-Rico2022,Fernandez-Rico2023,Meng2024,Vidal-Henriquez2021,Wei2020, Liu2023}.
However, unlike most prior 3D solid/liquid composite materials whose steady droplets are much larger than the characteristic mesh size of the polymer network \cite{Style2018,Rosowski2020,Rosowski2020a,Fernandez-Rico2022,Fernandez-Rico2023,Meng2024,Vidal-Henriquez2021,Wei2020}, the mesh size of the actin networks are typically 1-10 \textmu m, comparable to the lipid domain size on reconstituted membranes.
Thus, unlike prior studies on 3D solid/liquid composites, the 2D lipid domains do not experience continuum-scale interactions with the actin network, and the domains instead adopt highly irregular shapes as they conform to the network topology \cite{Arnold2023a, Arnold2023b, Gubbala2024, Honigmann2014, Vogel2017, LopesdosSantos2023}.
Moreover, due to the two-dimensionality of the system, even thin ($\approx$7 nm) actin filaments form impenetrable barriers that confine the domains without access to the third dimension \cite{Arnold2023b}.

\begin{figure}
\includegraphics[width=0.6\linewidth]{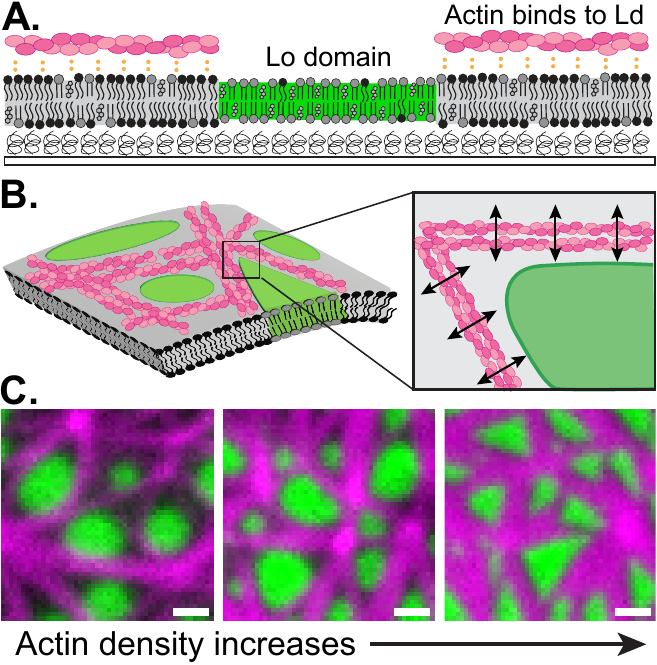}
\caption{
Actin-lipid membrane composite material design.
(A)
Side view of a reconstituted lipid membrane on a cushioned glass coverslip.
At room temperature, the membrane phase-separates into liquid-ordered (Lo, green) domains dispersed in a liquid-disordered (Ld, gray) continuous phase.
Filamentous actin (F-actin, magenta) is adsorbed to the Ld phase using electrostatic interactions.
(B)
(\emph{Left}) Orthographic projection of composite actin-coated lipid membranes.
(\emph{Right}) Top view of an actin junction near a Lo domain corner.
A lipid domain deforms such that the boundary is straighter along an actin bundle and more sharply curved near the intersection of multiple bundles.
(C)
Top-view micrographs of membranes with varying concentrations of actin (magenta) wetting the Ld phase while avoiding the Lo (green) phase.
Actin density increases from left to right, with domains developing sharper corners as actin density increases.
Scale bar is 1 \textmu m.
}
\label{Fig1_Overview}
\end{figure}

In this article, we present a composite 2D material in which liquid droplets control the viscoelasticity of an adsorbed actin network.
Elastic actin bundles are adsorbed to the liquid-disordered (Ld) continuous phase of a two-phase membrane without wetting liquid-ordered (Lo) dispersed domains (Fig. \ref{Fig1_Overview}A).
We recently showed that that active myosin-driven flows accelerate domain coarsening by 2$\times$ compared to thermal coarsening in such materials \cite{Arnold2023a}.
Here, we consider the two-way coupling between actin viscoelasticity and the Lo domain morphology and structure in the absence of myosin-driven activity. 
We find that the rigid actin bundles constrain the domains to adopt angular morphologies, deforming the Lo/Ld interface against line tension.
The domains, in turn, accelerate the viscoelastic relaxation of the actin network, decreasing the relaxation time $\tau$ by $\approx$2.5$\times$.
Finally, we rapidly change the chemical composition of the membrane, finding that the network stiffens as the domains grow, which provides a tool for controlling the mechanical properties of 2D composite interfaces.

\section{Experimental Methods}

\subsection{Buffers}
Filamentous actin buffer (F-buffer) consists of 50 mM Tris (pH 7.5), 2 mM magnesium chloride, 0.5 mM adenosine triphosphate (ATP), 0.2 mM calcium chloride, 25 mM potassium chloride, and 1 mM dithiothreitol (DTT). DTT is added to all buffers immediately before use to preserve its reactivity.
Assay buffer (A-buffer) consists of 25 mM imidazole (pH 7.4), 4 mM magnesium chloride, 1 mM (ethylene glycol-bis(\textbeta-aminoethyl ether)-N,N,N',N'-tetraacetic acid) (EGTA), 25 mM potassium chloride, and 1 mM DTT.
Globular actin buffer (G-buffer) consists of 2 mM Tris (pH 8.0), 0.2 mM calcium chloride, 0.5 mM DTT, 1 mM sodium azide, and 0.2 mM ATP.

\subsection{Actin and heavy meromyosin preparation}

Rabbit skeletal muscle actin is purified from muscle acetone powder (Pel-Freez, catalog no: 41995-2, Lot 16743) using standard methods \cite{Spudich1971, MacLean-Fletcher1980}.
No rabbits or other animals are directly involved in this study.
Actin is stored as depolymerized globular actin (G-actin) at -80\textdegree C in G-buffer with 6\% sucrose until use.

Actin is labeled with fluorescent Alexa Fluor 555 NHS Ester (Succinimidyl Ester) (Invitrogen catalog no: A20009) for microscopic visualization.
G-actin is reacted with NHS-Alexa Fluor 555 in HEPES buffer 
at room temperature for 30 minutes.
2x-concentrated F-buffer is then added to the G-actin so that Tris quenches the NHS reaction and F-buffer causes G-actin to polymerize to F-actin.
F-actin polymerization proceeds for 30 minutes at room temperature, and then overnight at 4\textdegree C.
Labeled F-actin is centrifuged at 142,000 $\times$ g for 30 minutes, and the pellet collected.
Unreacted dye and defective G-actin monomers and oligomers that are unable to polymerize are discarded in the supernatant.
The labeled F-actin pellet is washed with clean G-buffer, being careful not to disturb the pellet.
Labeled F-actin is dissolved in G-buffer, and allowed to de-polymerize for three days at 4\textdegree C before freezing and storing in 6\% sucrose at -80\textdegree C.

Heavy meromyosin (HMM) is purified from rabbit skeletal muscle (Pel-Freez Biologicals, Rogers, Arkansas) using standard methods \cite{Margossian1982}.
HMM is frozen in 10 mM potassium phosphate buffer (pH 7.0), 100 mM potassium chloride, 0.3 mM EGTA, 1 mM DTT, and 6\% sucrose at -80\textdegree C until use.

\subsection{Giant unilamellar vesicle (GUV) preparation}
Giant unilamellar vesicles (GUVs) are prepared using the established electroformation method \cite{Angelova1986}.
Briefly, lipids are mixed with the following composition: 45.7\% 1,2-dioleoyl-sn-glycero-3-phosphocholine (DOPC, Avanti catalog no: 850375P), 35\% 1,2-dipalmitoyl-sn-glycero-3-phosphocholine (DPPC, Avanti catalog no: 850355C), 15\% cholesterol (TCI Chemical, catalog no: C3624), 4\% 1,2-dioleoyl-3-trimethylammonium-propane (DOTAP, Avanti catalog no:	890890P), and 0.3\% 1,2-distearoyl-sn-glycero-3-phosphoethanolamine-N-[poly(ethylene glycol)2000-N'-carboxyfluorescein] (DSPE-PEG2k-FITC, Avanti catalog no: 	810120C).
This mixture of lipids and cholesterol phase-separates into liquid-ordered (Lo) domains dispersed within a liquid-disordered (Ld) continuous phase.
Lipids are spread on an indium tin oxide (ITO)-coated microscope slide (Diamond Coatings, 8-12 Ohm slide) and dried under vacuum for 30 minutes.

A 2 mm rubber gasket is sandwiched between the ITO-coated slide containing lipids, and a clean ITO-coated slide.
The interstitial space is then filled with 75 mM sucrose solution.
A sinusoidal electric potential of amplitude 3V (peak-to-peak) and frequency 10 Hz is applied to the chamber for one hour at 50 \textdegree C.
After one hour, the frequency is changed to 2 Hz for 30 minutes.
The resulting GUVs are collected, stored at room temperature and used within one day.

\subsection{Surface preparation}
Glass cover slips No. 1.5 (Azer Scientific) are cleaned with piranha solution (3:1 sulfuric acid:hydrogen peroxide) for five minutes and then washed with deionized water.
The cover slips are then made hydrophobic via reaction with trimethylchlorosilane (Sigma-Aldrich) vapors in a vacuum chamber, under house vacuum for ten minutes.
A 6 mm cylindrical polydimethylsiloxane (PDMS) chamber is attached to the cover slip surface to hold liquids.

Cover slips are incubated with 200 nM heavy meromyosin (HMM) for six minutes.
After six minutes, 0.1 mg/mL polylysine-grafted-PEG (PLL-g-PEG) is added and the HMM/PLL-g-PEG solution incubated for another three minutes.
The coverslip is then washed, first with A-buffer, and then with MilliQ water to remove unbound HMM and PLL-g-PEG.

This mixture of HMM and PLL-g-PEG on the coverslip surface helps to cushion the planar lipid bilayer and create space between the membrane and glass.
Lipid membrane domains experience frictional interactions with nanoscale roughness when directly supported on uncushioned glass coverslips.
Thus, without the cushion of bulky HMM and PEG, lipid domains would experience kinetic arrest and fail to grow beyond the diffraction limit \cite{Honigmann2012}.

\subsection{Adsorbing planar lipid bilayer to treated coverslip surface}
\label{sec:GUV_adsorption}
GUVs in MilliQ water are added to the actin/HMM-coated cover slip surface.
The cover slip is heated to 37\textdegree C for at least 15 minutes, during which time single-phase GUVs rupture on the treated surface, due to electrostatic attraction between positively-charged DOTAP and negatively-charged glass.
Unbound GUVs are then washed from the cover slip with A-buffer, leaving behind a planar, surface-adsorbed lipid bilayer.

\subsection{Assembling actomyosin cortex on a lipid bilayer}
Planar lipid bilayers adsorbed to cover slips are incubated in 1 \textmu M F-actin for 10 minutes at 37 \textdegree C.
F-actin adsorbs uniformly to the bilayer via electrostatic attraction to DOTAP \cite{Schroer2020,Heath2013}.
The membrane is then cooled to room temperature, causing it to phase-separate into liquid-ordered and liquid-disordered phases.
The actin then sequesters into the Ld phase.
Unbound actin is washed from the cover slip with A-buffer.
A complete phase-separated bilayer with actin adsorbed to the Ld phase is shown with uncropped edges in supporting Fig.~S1. 

\subsection{Inserting cholesterol into phase-separated membrane sample}
Cholesterol and methyl beta-cyclodextrin (m\textbeta CD) are combined in a 10:1 (chol:m\textbeta CD) ratio in MilliQ water and allowed to dissolve overnight at 50\textdegree C.
This mixture of cholesterol and m\textbeta CD is added to an actin/lipid bilayer sample at a working concentration of 5 \textmu M m\textbeta CD.
The m\textbeta CD solubilizes cholesterol in water and facilitates cholesterol incorporation into the lipid membrane.
Cholesterol incorporation alters the relative amounts of the Lo and Ld phases until the two phases eventually mix and form a single phase.
The membrane/actin sample is regularly imaged during this process.

\subsection{Preparing and tracking tracer actin}
Tracer actin filaments are prepared by polymerizing G-actin labeled with Alexa Fluor 555 in the presence of 100 nM gelsolin (Cytoskeleton, catalog no: HPG6-A \textit{Homo sapiens} recombinant).
Gelsolin truncates the resulting F-actin filaments, making them easier to track.
Dilute tracer F-actin is combined with full-length F-actin labeled with Alexa Fluor 405, such that the actin mixture contains 0.5\% mol G-actin-555.
This actin is then added to the lipid bilayer as described above.

Every few minutes, we capture a short series of images (100 total images, 0.2 s between images) of the tracer actin filaments
We track the movement of the tracer filaments using the Trackmate plugin in ImageJ \cite{Ershov2022}, and calculate the variance in tracking positions over the 20-second time lapse.

\subsection{Microscope for all imaging experiments}
All imaging is carried out on an inverted Nikon Ti2-Eclipse microscope (Nikon Instruments) using an oil-immersion objective (Apo 100x, NA 1.45, oil). Lumencor SpectraX Multi-Line LED Light Source is used for excitation (Lumencor, Inc). Fluorescent light is spectrally filtered with emission filters (432/36, 515/30, 595/31, and 680/42; Semrock, IDEX Health and Science) and imaged on a Photometrics Prime 95 CMOS Camera (Teledyne Photometrics). Microscope images are collected using MicroManager 1.4 software \cite{Edelstein2014}.

\subsection{Image analysis}
\subsubsection{Actin density}
To find actin density, we first binarize the Lo domain channel to eliminate areas of the membrane that are inaccessible to actin.
We calculate the area available to actin using $A_\mathrm{actin}=A_\mathrm{total}-A_\mathrm{Lo}$.
We then calculate the average fluorescence intensity of actin and divide it by the available area: $\rho=I_\mathrm{actin}/A_\mathrm{actin}$.

\subsubsection{Domain perimeter}
Domains are binarized using adaptive thresholding.
In many cases, binary thresholding merges tightly-spaced domains, which will greatly increase the excess perimeter if not corrected.
To separate merged domains, we take each binary domain of more than ten pixels in size and check whether it can be easily separated into multiple distinct objects.
First, the binarization threshold is increased until only 30\% of the original domain pixels remain.
The threshold is then iteratively lowered, 2\% at a time, and the number of distinct objects greater than 15 pixels in size in the resulting binary image counted.
If, after any threshold decrease, the number of binary objects in the image decreases, then the layer of pixels connecting the objects is deleted.
This process continues until the original threshold is recovered, effectively severing a single row of pixels between the domains.
The perimeter and area of each domain is then measured using the MATLAB Image Processing Toolbox.
Finally, we account for any remaining merged domains by removing any significant outliers in $L_\mathrm{ex}$, as determined by a Grubbs test.

\subsubsection{Domain curvature}
We take the binary domains, identified using the method described in the previous section, and fit a parametric curve to the boundary.
We initialize the parameter $t$ to have $5\times$ as many points as the number of edge pixels in the domain.
We fit a parametric curve with $\left(x\left( t\right) , y \left( t\right)\right)$ using smoothing splines in MATLAB.
We then calculate the unit normal vector $\mathbf{n}$ at each point in the curve, and find the curvature $\nabla \cdot \mathbf{n}$ using central finite differences.
We impose the fitted curve on a grid with the same spacing as the original image, and average the divergences that overlap with each grid point to produce the heatmap seen in Fig.~\ref{Fig2_Density}D.

\subsubsection{Relaxation time}
The relaxation time of the actin network is found using principles of differential dynamic microscopy (DDM), as described previously \cite{Cerbino2008,Giavazzi2009}.
The image structure function $D(\mathbf{k},t)$ is calculated by taking the fast Fourier transform of differences in image intensities $I$, separated by lag time $t$:
\begin{equation}
    D(k,t)=FFT\left[|I(\mathbf{x},t)-I(\mathbf{x},0)|^2\right].
    \label{eq:DDM}
\end{equation}
Here $k=\mathbf{k}$ is the magnitude of the wave vector, radially averaged over the 2D FFT output, and $\mathbf{x}$ is the spatial position vector.
In standard DDM analysis for ergodic systems, $D(k,t)$ is calculated via an ensemble average $\langle \cdot \rangle$ of Eq.~\ref{eq:DDM}.
However, as a lipid bilayer with adsorbed actin filaments, our material evolves in a non-ergodic fashion \cite{Weigel2011}, and we only consider one trajectory from initial to final state.

We find the intermediate scattering function $F(k,t)$ using the relation:
\begin{equation}
    F(k,t)=1-\frac{D(k,t)-B(k)}{A(k)}
\end{equation}
where $A(k)$ and $B(k)$ are time-independent constants.
We use the known asymptotic behavior $F(k,t\rightarrow \infty)=0$ and $F(k,t=0)=1$ to find $A(k)$ and $B(k)$, using the relations $A(k)+B(k)=D(k,t\rightarrow \infty)$ and $B(k)=D(k,t=0)$.
We find averaging the first and last few data points that qualitatively display asymptotic behavior to be the most effective way to find $F(k,t)$ in our system.
While $F(k,t)$ is exponential in time in most DDM experiments\cite{Cerbino2008}, our relaxation data is markedly non-exponential, thus making a two-parameter nonlinear fit for $A(k)$ and $B(k)$ inappropriate.
To avoid assuming a functional form for our relaxation data, we also define the relaxation time $\tau$ as the half life of the curve, such that $F(k,\tau)=0.5$.


\subsubsection{Domain and actin size}
We find the size of lipid domains and actin using the mean wave vector of the static structure factor $S(k)$.
To reduce artifacts due to photo bleaching, we first apply histogram-matched bleach correction to the time-lapse in ImageJ \cite{Miura2020}.
We take the 2D fast Fourier transform (FFT) of each image and take the absolute value of the Fourier coefficients to be $S(n)=|FFT(n)|$ where $n=0, 1, 2, ...$ is the Fourier mode.
We radially bin $S(n)$ to the nearest integer mode.
For example, the first diagonal point $S(n=\sqrt{2})$ would be rounded down to $n=1$ and averaged with all other coefficients $S(n=1)$.
Fourier modes are converted to wave vectors using the transformation $k=2\pi n/L$ where $L$ is the image width.
Only square images are analyzed, so that $L_x=L_y$, and thus $k_x=k_y$ for $n_x=n_y$.

All Fourier coefficients for high $k$, corresponding to wavelengths $2\pi/k$ of fewer than five pixels are discarded.
Due to photo bleaching, the domain intensity becomes smaller relative to the camera noise after long times (bleach correction cannot remove this artifact).
The camera noise is correlated across only a few pixels, so removing high wave vector data eliminates this artifact without affecting the more meaningful actin/domain data, which fluctuates along lower wave vectors.
We also remove the lowest wave vector, corresponding to the $n=0$ mode, as this mode corresponds only to an offset from zero intensity.
We find that removing $n=0$ does not change the trends in our data, but does result in more reasonable values of $\langle k \rangle$, as the $n=0$ coefficients typically dominate those of all higher modes.
We then find the mean wave vector $\langle k \rangle$ using Eq.~\ref{eq:kstar}.

\section{Results and Discussion}

\subsection{Membrane composite material design}

We achieve a planar, phase-separated lipid bilayer by rupturing a giant unilamellar vesicle (GUV) containing 46\% 1,2-dioleoyl-sn-glycero-3-phosphocholine (DOPC), 35\% dipalmitoylphosphatidylcholine (DPPC), 15\% cholesterol, 4\% 1,2-dioleoyl-3-trimethylammonium propane (DOTAP), and trace 1,2-distearoyl-sn-glycero-3-phosphoethanolamine-N-[poly(ethylene glycol)2000-N'-carboxyfluorescein] (DSPE-PEG2k-FITC, Lo-partitioning fluorescent dye) onto a polymer-cushioned cover slip (see Experimental Methods), causing the membrane to spread on the surface (supporting Fig.~S1).
We adsorb filamentous actin (F-actin) to the membrane at 37\textdegree C via electrostatic interactions between negatively-charged actin and positively-charged DOTAP \cite{Schroer2020, Heath2013}.
Upon cooling to room temperature, the membrane phase-separates, leaving Lo domains dispersed in a continuous Ld phase \cite{Veatch2003}.
Once the Lo domains nucleate amidst the actin network, they completely de-wet from actin, excluding it into the Ld phase (Fig.~\ref{Fig1_Overview}A-B, supporting Fig.~S1).

We previously observed this near-total partitioning of actin into the Ld phase in similar experiments \cite{Arnold2023a,Arnold2023b}, and expect it to be caused in part by the imbalance of charge between the two phases, as DOTAP strongly partitions into the Ld phase \cite{Honigmann2014}.
However, simulations have shown that when actin adsorbs electrostatically to a lipid bilayer, the bilayer deforms out-of-plane to wet the actin \cite{Schroer2020}.
Thus, we expect that the stiffer Lo phase also resists the necessary bending to wet actin, ensuring near-total partitioning into the Ld phase.

The fluorescence micrographs in Fig.~\ref{Fig1_Overview}C suggest that the thin $\sim \mathcal{O}$(nm) actin filaments align and cluster to form thick $\sim \mathcal{O}$(\textmu m) bundles, as has previously been observed in electrostatically-adsorbed actin on lipid membranes \cite{Heath2013}.
We observe that when small amounts of actin are adsorbed to the membrane surface, the thin bundles assemble into a disordered network that avoids and deforms around the domains (Fig.~\ref{Fig1_Overview}C, left panel).
However, if more actin is allowed to adsorb, the bundles thicken and ultimately span nearly the entire Ld phase (Fig.~\ref{Fig1_Overview}C, center and right panels).
These thicker bundles interact with the Lo/Ld interface, straightening the domain edges next to and along the bundles, while sharply deforming the domain edges near the intersections of actin bundles (Fig.~\ref{Fig1_Overview}B-C).

\subsection{Actin elasticity heterogeneously deforms domains}

\begin{figure}
\includegraphics[width=\linewidth]{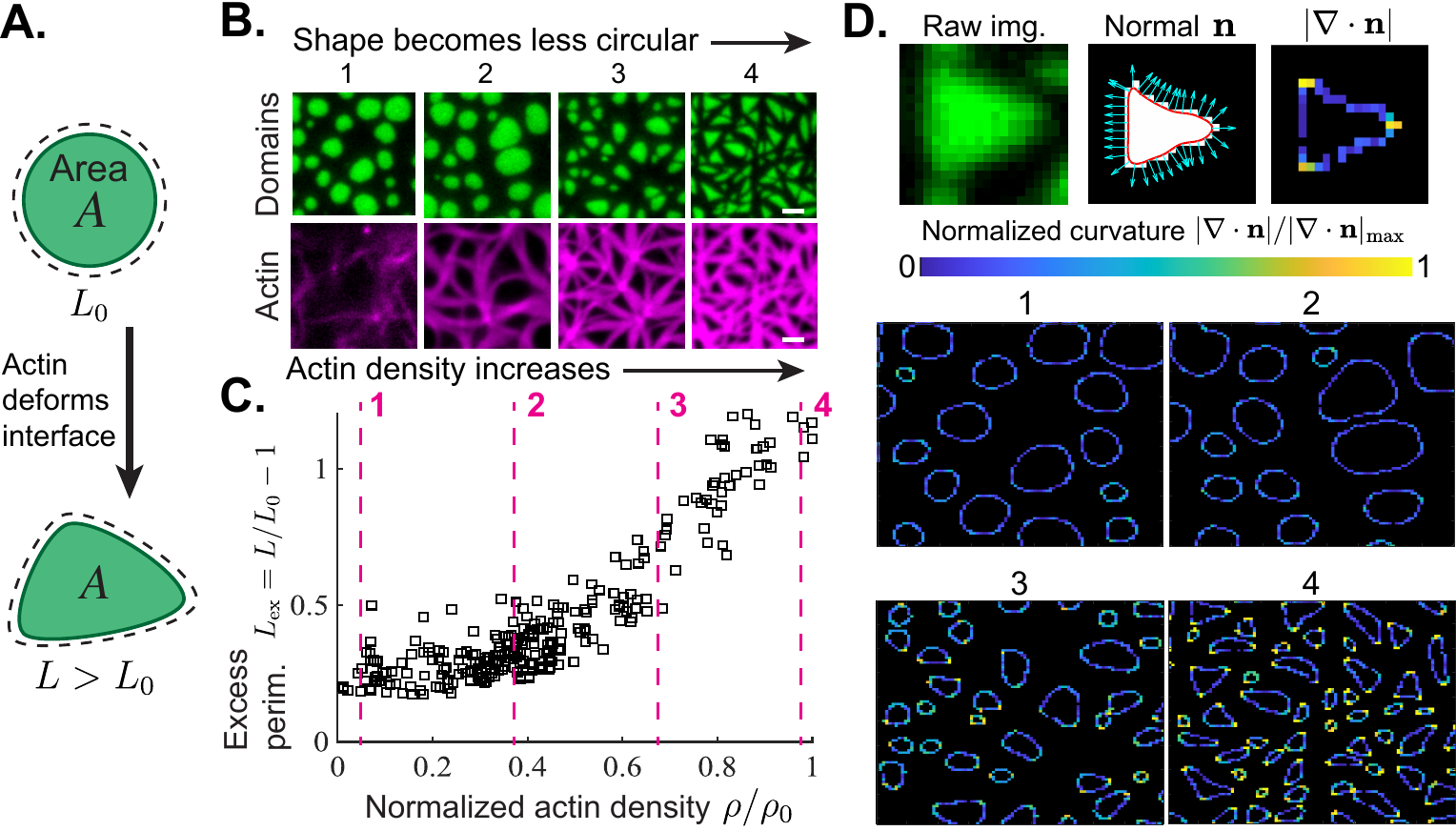}
\caption{
High actin density deforms lipid domain boundaries, increasing the perimeter and anisotropy.
(A)
In its relaxed state, a liquid ordered (Lo) lipid domain of area $A$ adopts a circular morphology with perimeter $L_0$.
Interactions with the actin network deforms the domain and its interfacial perimeter increases to $L>L_0$.
(B)
(\emph{Top row}) Top-view micrographs of a phase-separated lipid membrane with Lo domains in green and the liquid-disordered (Ld) continuous phase in black.
(\emph{Bottom row}) Actin filaments (magenta) are adsorbed to the Ld phase of the lipid bilayer.
As actin density increases from left to right, the lipid domains change from a circular morphology to a triangular one.
Scale bars are 2.5 \textmu m.
(C)
Excess perimeter $L_\mathrm{ex} \equiv (L-L_0)/L_0$ of lipid domains is plotted as a function of actin density.
Actin density $\rho$ represents mean fluorescence intensity of actin, divided by the available area of the Ld phase, and is normalized by the maximum value sampled $\rho_0$.
Pink dashed lines with numbers 1-4 correspond to the actin densities in the images in Fig.~2B.
Scatter plot includes n=270 independent lipid bilayer/actin samples.
(D)
(\emph{Top})
Local domain curvatures are calculated for images 1-4 in Fig.~2B.
A raw image of each domain is converted to a binary image, and a parametric curve fit to the domain boundary using smoothing splines.
The curvature $\nabla \cdot \mathbf{n}$ is calculated at every point on the fitted curve using the unit normal $\mathbf{n}$ (cyan arrows).
(\emph{Bottom})
Heat maps show the large spatial heterogeneity in the local curvature for images 1-4, normalized by the maximum value in image 4, $|\nabla \cdot \mathbf{n}|/|\nabla \cdot \mathbf{n}|_\mathrm{max}$.
Black pixels indicate zero curvature.
}
\label{Fig2_Density}
\end{figure}

In its relaxed state, a lipid domain of area $A$ minimizes its perimeter by adopting a circular shape with perimeter $L_0=2\sqrt{\pi A}$.
Deforming a domain of constant area into a non-circular shape of perimeter $L>L_0$ increases the free energy by $\Delta F =\lambda(L-L_0)$ where $\lambda$ is the line tension of the Lo/Ld interface (Fig.~\ref{Fig2_Density}A).
We define the excess perimeter $L_\mathrm{ex}$ to be the difference between the domain perimeter and the minimum perimeter of a circle with equivalent area
\begin{equation}
    L_\mathrm{ex}\equiv \frac{L-L_0}{L_0}=\frac{L}{2\sqrt{\pi A}}-1 .
    \label{eq:ExcessPerim}
\end{equation}
In Fig.~\ref{Fig2_Density}B-C, as the density of adsorbed actin in the Ld phase increases, the excess perimeter of lipid domains increases, corresponding to a shift from circular to triangular morphologies.

We hypothesize that the mechanism by which actin deforms the Lo/Ld domain interface relies on actin's bending rigidity and its tendency to nematically align on the lipid membrane.
Heath et al. showed that actin filaments adsorbed electrostatically to Ld-phase supported lipid bilayers will align into a liquid crystalline structure with quasi-long ranged nematic order and periodic filament spacing \cite{Heath2013}.
We polymerize globular actin (G-actin) into F-actin ex situ, and then adsorb F-actin to a heated, single-phase membrane.
Thus, filaments bind in all orientations, and then rotate to form parallel bundles on the fluid surface, subject to limitations due to entanglements and crossovers between filaments.

As the system cools and ages, Lo domains nucleate and grow, excluding the actin filaments into a shrinking Ld continuous phase.
We note that actin does not directly contact the Lo domains, but rather is adsorbed atop the Ld phase.
Thus domains cannot pass under actin filaments, but individual lipids from the Lo phase can dissolve into and diffuse through the Ld phase.
As Lo domains compress actin into a shrinking Ld area, actin filaments must rotate and diffuse so as to relax steric and electrostatic repulsion between filaments, increasing nematic alignment of bundles.
However, the Lo domains simultaneously frustrate actin alignment and relaxation because they are effectively incompressible, 2D droplets.
Actin bundles separated by Lo domains cannot merge unless the intervening domains completely dissolve into the Ld phase, clearing a path between them.
Relaxation is further limited when bundles cross one another and entangle, as seen in the bright magenta spots in Figs.~\ref{Fig1_Overview}C and \ref{Fig2_Density}B.

Rigid, kinetically-frustrated actin bundles deform Lo domains into highly anisotropic shapes and increase the interfacial free energy of the material.
Actin has a flexural rigidity $EI=2\times10^{-25}$ N$\cdot \mathrm{m}^2$ \cite{Jacobs2012}, meaning the force necessary to buckle an actin filament of length $L=1$ \textmu m is $F_\mathrm{b}=\pi EI/L^2=$ 1 pN.
Phase-separated lipid domains have line tension $\lambda = 1-10$ pN \cite{Tian2007}, suggesting that small bundles of only a few $1-10$ \textmu m filaments will bend around an Lo domain, minimally disturbing the circular interface (Figs.~\ref{Fig1_Overview}C and \ref{Fig2_Density}B, left panels).
Assuming $\approx$ 20 nm spacing between filament centers \cite{Heath2013}, we estimate that the $\approx$ 1-2 \textmu m actin bundles are $\approx$ 50-100 filaments thick.
Thus at higher actin densities, bundle stiffness dominates line tension, and the domains deform to accommodate the actin bundles (Fig.~\ref{Fig2_Density}C).
We note that while the domains are sharply deformed at high actin densities, their total area appears approximately constant, suggesting that while actin filaments can deform domain shape, they are not adsorbed to the Ld phase strongly enough to suppress phase separation.

Previously, micron-sized liquid inclusions were grown in 3D composite materials with nanometer-scale spacing between elastic polymers \cite{Style2015,Style2018,Rosowski2020,Rosowski2020a,Fernandez-Rico2023}.
Thus, the droplets experienced effectively an isotropic elastic stress exerted by a continuum polymer gel.
In our 2D material, the actin bundles and Lo domains are both $\sim \mathcal{O}$(\textmu m) in size, causing the domains to deform anisotropically in response to actin elasticity.
To characterize the anisotropic elastic stresses on domains, we fit parametric curves to the domain boundaries, find the unit normal $\mathbf{n}$, and calculate the local curvature $\nabla \cdot \mathbf{n}$ along the boundary (Fig.~\ref{Fig2_Density}D).
At low actin densities, actin exerts little influence on domain shape, resulting in relatively uniform curvature (Fig.~\ref{Fig2_Density}D, panels 1 and 2).
Meanwhile, dense actin bundles elicit large heterogeneity in curvature within each domain boundary, with high curvature concentrating in only a few pointed regions (Fig.~\ref{Fig2_Density}D, panels 3 and 4).

In a quasi-static material with no flows, the curvature can be connected to the elastic stress using a modified Laplace-Young equation:
\begin{equation}
    P_\mathrm{Lo}-P_\mathrm{Ld} = \lambda \nabla \cdot \mathbf{n} + \sigma_\mathrm{el} .
    \label{eq:LaplaceYoung}
\end{equation}
The surface pressures in the Ld ($P_\mathrm{Ld}$) and Lo ($P_\mathrm{Lo}$) phases have units of force/length, and are balanced by line tension $\lambda$ and a compressive elastic stress due to actin $\sigma_\mathrm{el}$.
While equations of this form are often used in 3D composites \cite{Meng2024, Wei2020, Vidal-Henriquez2021}, our Lo domains do not contact actin directly, but rather experience elastic stress transmitted through the Ld phase.
Thus, we combine $\sigma_\mathrm{el}$ and $P_\mathrm{out}$ into an effective outside pressure, $P_\mathrm{Ld}^\mathrm{eff}= \sigma_\mathrm{el} + P_\mathrm{out}$, which simplifies Eq.~\ref{eq:LaplaceYoung}: $P_\mathrm{Lo}-P_\mathrm{Ld}^\mathrm{eff} = \lambda \nabla \cdot \mathbf{n}$.

In a quasi-static membrane, lipid flows within the purely viscous Lo phase are assumed to be minimal, suggesting that $P_\mathrm{Lo}$ is spatially invariant within each Lo domain.
Thus, the spatial variations in $\nabla \cdot \mathbf{n}$ seen in Fig.~\ref{Fig2_Density}D are balanced by variations in $P_\mathrm{Ld}^\mathrm{eff}$, which arise due to the actin elasticity compressing the domain anisotropically.
Previous work has focused only on heterogeneities in matrix stiffness between 3D droplets, with each droplet still experiencing isotropic compression \cite{Meng2024}.
We hypothesize that the two-dimensionality of our material, which encourages nematic actin bundling and thus maintains parity between the network and droplet length scales, is responsible for this unusual anisotropic stress concentration.
Thus, the actin elasticity dictates the lipid domain architecture, and actin density provides a mechanism of controlling droplet morphology.

\begin{figure}
\includegraphics[width=\linewidth]{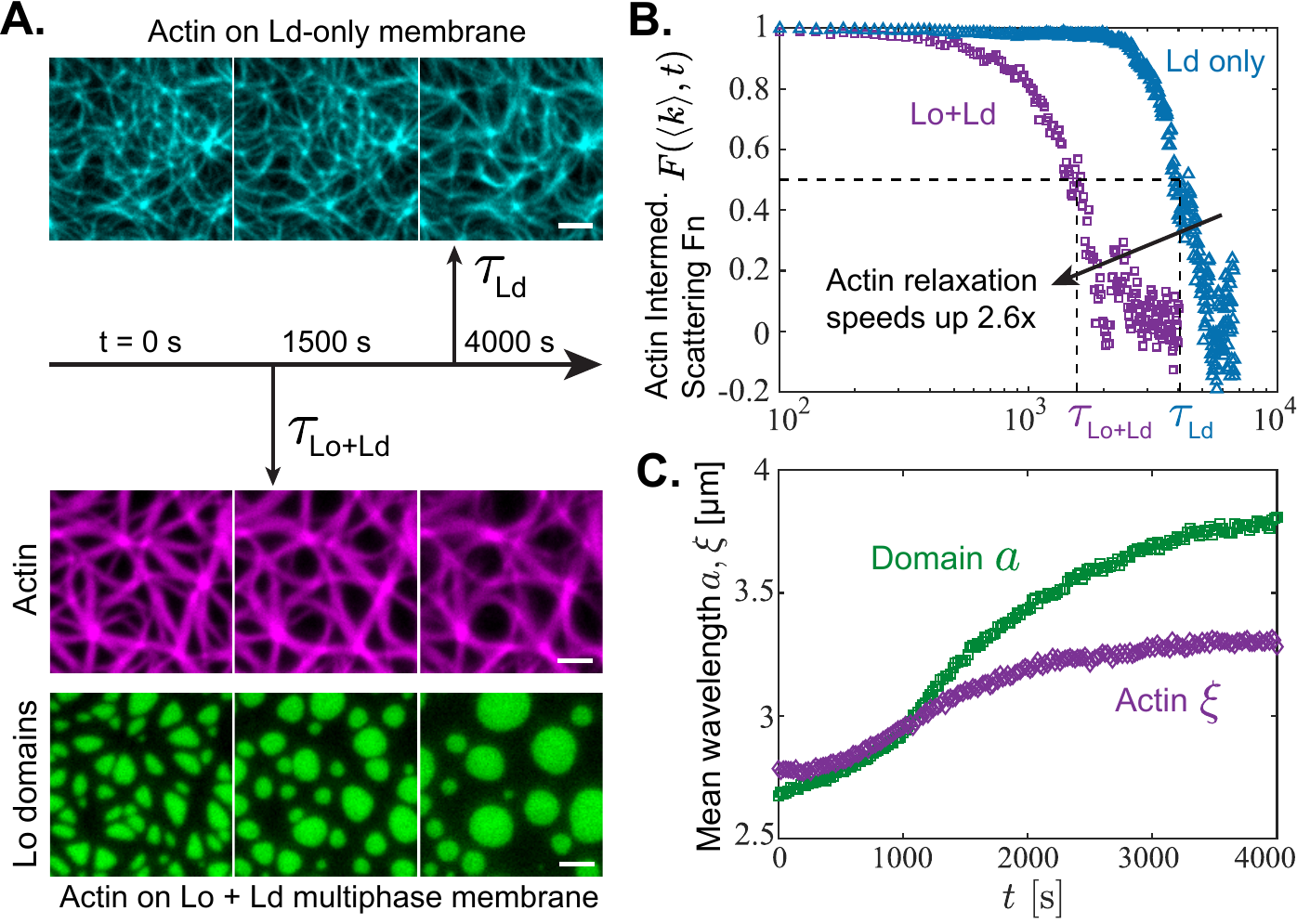}
\caption{
Coarsening lipid domains accelerate the relaxation and ageing of viscoelastic actin networks.
(A)
(\emph{Top}) Fluorescence micrographs show actin (cyan) on a single-phase liquid-disordered (Ld) membrane during a 4000 s time-lapse.
(\emph{Bottom})
A separate sample containing actin (magenta) and liquid-ordered (Lo) domains (green) on a two-phase Ld/Lo membrane is also imaged for 4000 s.
Relaxation time of the actin network on the two-phase Ld/Lo membrane ($\tau_\text{Lo+Ld}$) was 2.6$\times$ smaller than that on the single-phase Ld membrane ($\tau_\text{Ld}$).
Scale bars are 2.5 \textmu m.
(B)
The intermediate scattering function $F(k,t)$ is plotted as a function of time $t$ for the mean wave vector $k=\langle k \rangle$ of each sample (see Eq.~\ref{eq:kstar} for definition of $\langle k \rangle$), for both the two-phase (Lo+Ld) and one-phase (Ld only) samples.
The relaxation time $\tau$ for each sample is the half life, $F(\langle k \rangle,\tau)=0.5$. 
(C)
The mean wavelength of the static structure factor, defined as $a,\xi = 2\pi/\langle k \rangle$, is plotted for both domains $a$ and the actin network $\xi$.
Actin and domains both follow the same qualitative trend during coarsening, with the coarsening rate initially accelerating, before slowing after $\approx$1500 s.
}
\label{Fig3_Time-lapse}
\end{figure}

\subsection{Coarsening lipid domains accelerate the relaxation of viscoelastic actin networks}

Actin and lipid domains are two-way coupled: actin controls the domain morphology while the domains themselves control actin viscoelasticity. 
While the actin network structure remains quasi-static over short time scales of seconds to minutes, the structure slowly relaxes over tens of minutes to hours.
In Fig.~\ref{Fig3_Time-lapse}A, we consider the actin network structure on both a single-phase Ld membrane (top) and a two-phase Lo/Ld membrane (bottom).
Due to the high variability in actin adsorption to the membrane between trials, we present a single representative pair of membranes with similar actin densities.
We use differential dynamic microscopy (DDM) to calculate the intermediate scattering function $F(k,t)$ of actin for wave vector magnitude $k=|\mathbf{k}|$ over time $t$ (see Methods) \cite{Cerbino2008,Giavazzi2009}.
We define the mean wave vector $\langle k \rangle$ of actin in each sample using the static structure factor $S(k)$
\begin{equation}
    \langle k \rangle=\frac{\int kS(k) \mathrm{d}k}{\int S(k) \mathrm{d}k}
    \label{eq:kstar}
\end{equation}
and plot $F(\langle k \rangle, t)$ in Fig.~\ref{Fig3_Time-lapse}B.

Actin network evolves slowly on a single-phase, Ld-only membrane, with a relaxation time $\tau_{\text{Ld}} \approx 4100$ s, where $\tau$ is defined as the half life of the intermediate scattering function: $F(k,\tau)=0.5$.
However, the actin network relaxes more quickly on a two-phase Lo/Ld membrane of similar overall actin density, with $\tau_{\text{Lo+Ld}} \approx 1600$~s (Fig.~\ref{Fig3_Time-lapse}B, supporting Video~S1).
While $\langle k\rangle$ is $\approx$1 \textmu m\textsuperscript{-1} smaller in the single-phase system, the two-phase membrane relaxes faster than the single-phase across many wave vectors (supporting Fig.~S3).
Actin responds elastically to deformations, but is free to diffuse along the 2D fluid membrane, giving the composite membranes with adsorbed actin viscoelastic properties.
The $\approx$2.6$\times$ acceleration in actin structural relaxation upon adding Lo inclusions to the membrane demonstrates that the domains control the viscoelastic behavior of the material.

This is significant because the relaxation time is a readout of the elastic to viscous crossover of a material.
Thus, in reducing the relaxation time, the lipid inclusions accelerate the onset of the viscous behavior of the actin network, making the network effectively softer over these intermediate time scales.
This is a key difference with prior work on large 3D droplets immersed in a small-molecule polymer network \cite{Style2015}, where the droplets stiffened the overall composite material.

The images in Fig.~\ref{Fig3_Time-lapse}A offer a qualitative explanation of the changing relaxation dynamics: the lipid domains coarsen amongst the actin network, providing a driving force for relaxation.
At early times, we observe coarsening primarily via Ostwald ripening: lipids from small Lo domains dissolve in the Ld phase, and then re-condense in larger Lo domains (supporting Video~S2).
The large domains exclude actin as they grow, while the small domains free up space for actin as they dissolve.
At very long times, actin appears to partially unbind from the surface, giving Lo domains more freedom to coalesce (Fig.~\ref{Fig3_Time-lapse}A, bottom, supporting Video~S2).

The coupling between actin relaxation and lipid domain coarsening becomes apparent when we evolve the mean wavelength of the structure factor $a, \xi = 2\pi/\langle k\rangle$ of both domains $a$ and actin $\xi$ (Fig.~\ref{Fig3_Time-lapse}C).
For the first $\approx$1500~s, $a$ and $\xi$ nearly overlap, with both growing increasingly quickly as the actin network relaxes and domains coarsen.
This acceleration in growth is significant because passive lipid membrane domains grow according to a power law $a\sim t^{1/3}$ with sub-linear scaling \cite{Lifshitz1961,Camley2011}.
Elasticity reduces this scaling in 3D systems as it resists droplet deformation \cite{Meng2024}.

Here, the growing lipid domains are initially frustrated by actin elasticity, but at longer times, the actin relaxes in response to sustained deformation from the coarsening domains.
Similar viscoelastic relaxation has been observed in 3D fibrillar networks, in which coarsening droplets drive sporadic and abrupt plastic deformations in the network architecture \cite{Liu2023}.
At long times, the actin begins to unbind from the membrane, causing $\xi$ to plateau around $\approx$2000~s, while the domains continue to coarsen (Fig.~\ref{Fig3_Time-lapse}C).
The similar unbinding we see in the single-phase case at $\approx$4000 s (supporting Video~S1) confirms that viscoelastic 2D actin networks will always unbind from membranes at long times, but that the domains drive these dynamics much more rapidly.

\begin{figure}
\includegraphics[width=\linewidth]{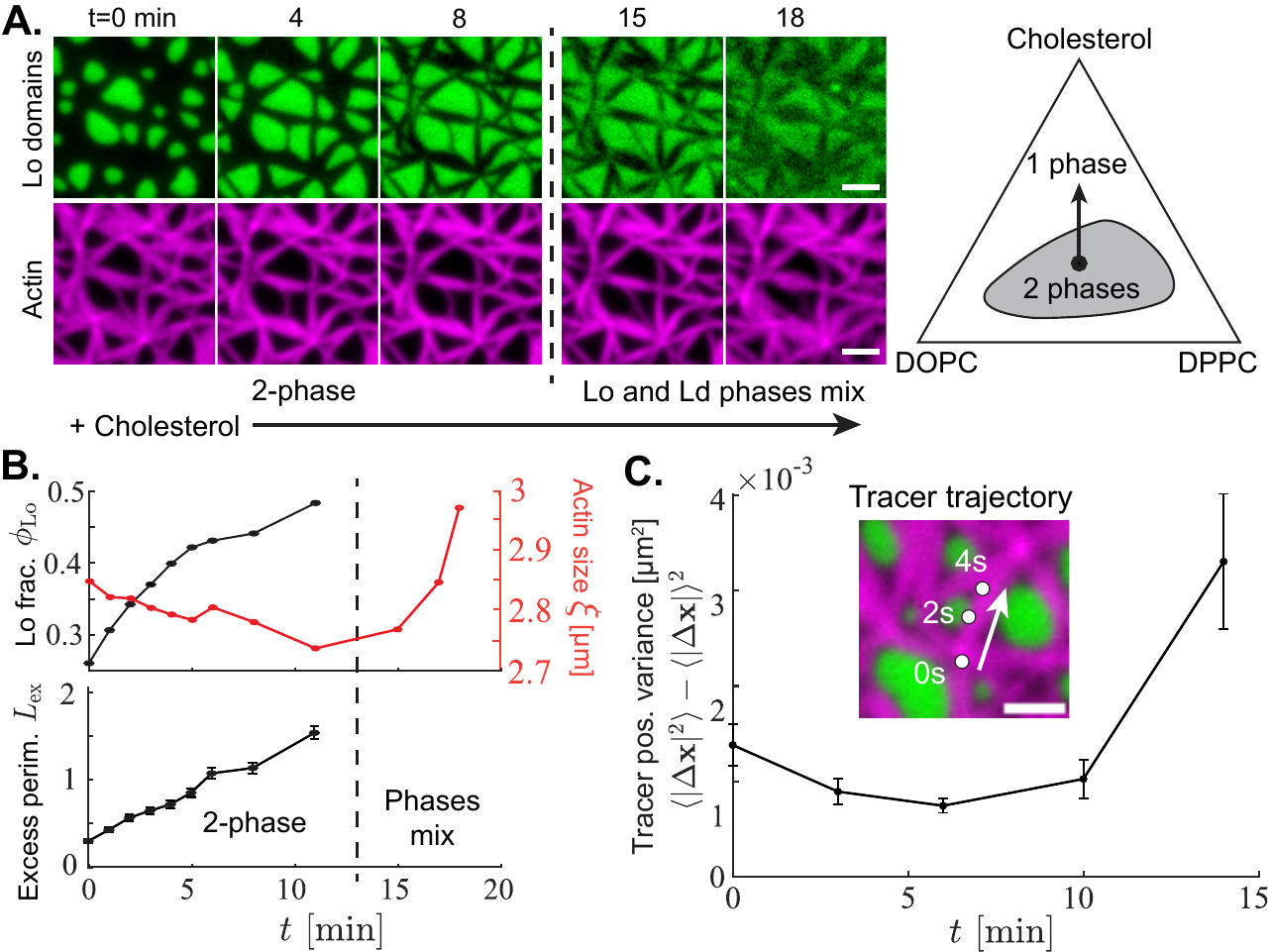}
\caption{
Exogenous cholesterol addition shifts membrane phase equilibrium, changing the mechanical properties of the composite material.
(A)
Fluorescence micrographs (left) show actin (magenta) adsorbed to a phase-separated lipid membrane with liquid-ordered (Lo, green) domains coexisting with a liquid-disordered (Ld, black) continuous phase.
Cholesterol is added to the membrane beginning at time $t=0$.
The ternary phase diagram (right) for a membrane with DOPC, DPPC, and cholesterol shows that the fraction of DPPC-rich Lo phase initially increases while the membrane composition remains in the two-phase region.
Eventually, the membrane crosses the binodal curve and enters the one-phase region.
This phase change occurs between the 8 and 15 minute snapshots on the left.
Scale bars are 2.5 \textmu m.
(B)
The Lo domain area fraction $\phi_\mathrm{Lo}$, actin mean wavelength $\xi$, and domain excess perimeter $L_\mathrm{ex}$ are plotted as a function of time.
After the 11 minute measurements, the Lo and Ld phases begin to mix, so the domain properties $L_\mathrm{ex}$ and $\phi_\mathrm{Lo}$ become undefined.
(C)
Dilute tracer filaments are embedded in a more dense actin network, but are labeled with a different fluorescent dye.
Over short ($<$10 s) intervals, the variance in tracer positions $\langle |\Delta \mathbf{x}|^2 \rangle - \langle |\Delta \mathbf{x}| \rangle^2 $ is tracked.
The mean of n=212 tracer variances is plotted against time since cholesterol addition, with error bars representing the standard error of the mean.
Inset shows the position of a tracer amongst the actin network over a period of 6 s.
Scale bar is 2.5 \textmu m.
}
\label{Fig4_chol}
\end{figure}

\subsection{Dynamic modulation of membrane composition drives a non-monotonic shift in actin stiffness}

We have shown that domain coarsening drives actin relaxation, driven by the thermodynamic driving force of domains to grow and displace actin.
However, if the membrane composition is perturbed so that the Lo area fraction increases, then the actin bundles may be constrained into an even smaller area, resulting in network stiffening.
We dynamically control membrane composition by introducing methyl beta-cyclodextrin (m\textbeta CD)-solubilized cholesterol to the membrane, so that cholesterol incorporates into the inner leaflet \cite{Lee2020,Ohtani1989,Klein1995,Roper2000}.

We observe that the Lo domains grow larger initially because cholesterol inserts preferentially into the Lo phase, consistent with the ternary phase diagram for DOPC, DPPC, and cholesterol developed by Veatch and Keller (Fig.~\ref{Fig4_chol}A) \cite{Veatch2003}.
However, after $\approx$15 minutes, the cholesterol concentration exceeds the binodal curve in the phase diagram, and the two phases begin to mix (Fig.~\ref{Fig4_chol}A).
We believe that the strong contacts with actin reduce lipid mobility, and are responsible for the relatively slow mixing time scale.

During the initial period of Lo domain growth (first $\approx$11 min), the area fraction of Lo phase $\phi_\mathrm{Lo}$ increases from approximately 20\% to almost 50\% (Fig.~\ref{Fig4_chol}B).
This compresses the Ld-tethered actin into a much smaller space, narrowing actin bundles and decreasing the mean wavelength $\xi$ (red curve in Fig.~\ref{Fig4_chol}B).
As actin is more tightly compressed, its increased elastic forces resist further deformation by the growing Lo domains.
Thus, the domains deform sharply to fill all available space, increasing $L_\mathrm{ex}$ as sharp corners form (Fig.~\ref{Fig4_chol}B).
The dominance of actin elasticity over line tension is magnified by the increasing cholesterol composition, which reduces line tension \cite{Tsai2019} and allows $L_\mathrm{ex}$ to grow beyond that of any sample in Fig.~\ref{Fig2_Density}.

As the actin bundles are compressed by the growing domains, they stiffen and suppress thermal fluctuations of individual filaments.
We measure the fluctuations of dilute, short ``tracer'' actin filaments, which are labeled with a different fluorescent dye so that they can be individually tracked amongst the bulk actin (supporting Fig.~S4).
The tracers are also truncated with the actin-severing protein gelsolin, to reduce friction and entanglements with other filaments, and to ease tracking (supporting Fig.~S4).
We measure the variance in position $\langle|\Delta \mathbf{x}|^2\rangle - \langle|\Delta \mathbf{x}|\rangle^2$ of n=212-254 tracer filaments over short, 20-second time-lapses, taken every few minutes during cholesterol addition (Fig.~\ref{Fig4_chol}C, supporting Video~S3).

The variance of tracer positions initially decreases as cholesterol is added to the membrane.
As the actin bundles are compressed into a smaller area, crowding and friction between filaments limit tracer Brownian motion.
Upon adding sufficient cholesterol, the composition exceeds the binodal of the phase diagram (Fig.~\ref{Fig4_chol}A) and the domains mix to form a single liquid phase.
As the actin-free Lo areas disappear, the actin bundles begin to spread out to occupy the entire membrane, with $\xi$ increasing well beyond its initial value (Fig.~\ref{Fig4_chol}B).
Accordingly, the tracer actin filaments become considerably more mobile as they gain access to an additional $\approx$25\% of the membrane, reducing friction and crowding (Fig.~\ref{Fig4_chol}C).

These tracer mobility trends suggest that Lo domains stiffen the 2D composite material, seemingly contradicting our earlier results of network softening via coarsening lipid domains.
However, there are key differences between changing the molecular composition of the membrane and allowing domains of constant composition to nucleate and coarsen.
In a coarsening membrane, the area fractions of Lo and Ld are constant, and large Lo domains grow by consuming smaller domains.
The growing domains exclude actin at the same rate that the shrinking or consumed domains accommodate it.
Thus actin is effectively mixed by the mass-conserved transport of lipids and cholesterol.

Cholesterol addition rapidly grows all Lo domains, large and small, while simultaneously reducing line tension and thus slowing coarsening.
The resulting domain growth dominates coarsening, locking actin bundles into their original network structure.
As the networks are compressed by the shrinking Ld phase, the material stiffens.
An analogous 3D system might have liquid inclusions that dynamically hydrate or dehydrate the surrounding polymer gel, increasing or decreasing its plasticity.
Modulating the cholesterol composition provides a mechanism to control the viscoelasticity of our 2D material.

\section{Conclusions}
We present a 2D composite material with lipid domains embedded in a viscoelastic actin matrix, and use these liquid inclusions the tune the viscoelasticity of the material.
Unlike prior work on 3D composite gels \cite{Style2015,Style2018,Rosowski2020,Rosowski2020a}, the characteristic sizes of the actin network and lipid domains are similar, resulting in anisotropic stresses that deform lipid domains away from a circular shape.
The 2D nature of our material enhances this effect, as it aligns the actin filaments and prevents domains and filaments from ``escaping'' into the third dimension to dissipate their stress.

Not only do the actin bundles impose anisotropic morphologies on the domains, but the domains control the viscoelasticity of the actin network.
Domain coarsening accelerates actin relaxation, driving an earlier onset of viscous behavior, which effectively softens the material at intermediate shear frequencies.
We dynamically control the viscoelasticity of the network by inserting cholesterol into the membrane, manipulating the lipid membrane composition to stiffen or soften the material.
These 2D liquid inclusions thus provide a mechanism through which to both stiffen and soften viscoelastic interfaces.

Mammalian cells achieve exceptional plasticity when crawling through complex environments \cite{Phillips2009}, all while constantly deforming, stretching, and organizing their complex plasma membrane.
Reconstituted solid/liquid composite membrane materials recapture some of this complexity, and offer the possibility to develop novel interfacial materials with spatiotemporally-tunable viscoelastic properties \cite{Arnold2023b}.

\begin{suppinfo}

Videos showing actin relaxation on both single- and two-phase membranes, actin relaxation with simultaneous domain coarsening, and tracer actin fluctuation; note providing additional details on cholesterol insertion into membrane; figure showing margins of composite membrane material; and figure showing tracer actin with different fluorescent dyes.

\end{suppinfo}

\begin{acknowledgement}
The authors thank Aakanksha Gubbala for helpful discussions regarding the theory of phase-separation in an elastic medium.
This material is based upon work supported by the National Science Foundation under Grant No.~2150686.
D.P.A. is supported by the National Science Foundation Graduate Research Fellowship under Grant No.~2139319.
S.C.T. is supported by the Packard Fellowship in Science and Engineering.
\end{acknowledgement}

\providecommand{\latin}[1]{#1}
\makeatletter
\providecommand{\doi}
  {\begingroup\let\do\@makeother\dospecials
  \catcode`\{=1 \catcode`\}=2 \doi@aux}
\providecommand{\doi@aux}[1]{\endgroup\texttt{#1}}
\makeatother
\providecommand*\mcitethebibliography{\thebibliography}
\csname @ifundefined\endcsname{endmcitethebibliography}  {\let\endmcitethebibliography\endthebibliography}{}

\end{document}